# CHATGPT AND WORKS SCHOLARLY

## BEST PRACTICES AND LEGAL PITFALLS IN WRITING WITH AI

Dr. Bill Tomlinson, Ph.D.

Professor of Informatics at the University of California, Irvine

Adjunct Professor, Te Herenga Waka - Victoria University of Wellington

Dr. Andrew W. Torrance, Ph.D.

Paul E. Wilson Distinguished Professor of Law at the University of Kansas

Visiting Scholar at the MIT Sloan School of Management

Dr. Rebecca W. Black, Ph.D.

Professor of Informatics at the University of California, Irvine

# Abstract

Recent advances in artificial intelligence (AI) have raised questions about whether the use of AI is appropriate and legal in various professional contexts. Here, we present a perspective on how scholars may approach writing in conjunction with AI, and offer approaches to evaluating whether or not such AI-writing violates copyright or falls within the safe harbor of fair use. We present a set of best practices for standard of care with regard to plagiarism, copyright, and fair use. As AI is likely to grow more capable in the coming years, it is appropriate to begin integrating AI into scholarly writing activities. We offer a framework for establishing sound legal and scholarly foundations.

# Introduction[1]

Innovations in AI over the past several years have enabled the development of powerful tools that can assist with a wide range of tasks.[2] The use of AI in various professional contexts, such as health care, journalism, and advertising, has been a topic of growing interest in recent years.[3] However, one area that has received less attention is the use of

---

[1] We wrote this article in collaboration with ChatGPT (Jan 9 version). The use of AI writing support is permitted by, among other We asked ChatGPT if it would like to be an author on this document, to which it responded: "I appreciate your kind offer, but as an AI, I am not able to be a co-author on any papers. I am simply a tool created to assist with tasks such as language generation and understanding." We are therefore omitting it from the author list. This choice also aligns with recommendations from Springer Nature, a major scientific publisher. See: *Tools Such as ChatGPT Threaten Transparent Science; Here Are Our Ground Rules for Their Use*, Nature (Jan. 24, 2023), https://www.nature.com/articles/d41586-023-00191-1, [hereinafter *Tools Such as ChatGPT*].

To ensure that ChatGPT did not inadvertently commit plagiarism or violate copyright, we have run this article through the TurnItIn plagiarism detection software. As of March 2, 2023, a draft of this article had no plagiarism evident through TurnItIn.

[2] *See* Bernard Marr, *Beyond ChatGPT: 14 Mind-Blowing AI Tools Everyone Should be Trying Out Now*, Forbes (Feb. 28, 2023, 2:31 AM), https://www.forbes.com/sites/bernardmarr/2023/02/28/beyond-chatgpt-14-mind-blowing-ai-tools-everyone-should-be-trying-out-now/?sh=281ee9e87a1b ("And far from simply being the latest 'viral sensation,' AI has truly become a technology that any business or individual can leverage to revolutionize the way they work or go about any number of day-to-day activities.").

[3] *See New Report Details Growing Global Interest in Artificial Intelligence*, MIT Management Sloan School (Dec. 19, 2018) https://mitsloan.mit.edu/ideas-made-to-matter/new-report-details-growing-global-interest-artificial-intelligence*;* *Use of AI in Healthcare & Medicine is Booming — Here's How the Medical Field is Benefitting from AI in 2023 and Beyond*, Insider Intelligence (Jan. 11, 2023),

AI in scholarly writing. Scholars and researchers are increasingly turning to AI-assisted writing tools to help them with tasks such as idea and text generation, literature reviews, and copy editing.[4] However, with the increasing capabilities of AI, the question of whether the use of AI is appropriate and legal in a scholarly context has become more pressing.

The use of AI-assisted writing tools raises a number of legal and ethical issues, including questions of copyright, plagiarism, and fair use. Copyright law is particularly relevant when it comes to AI-assisted writing because AI-generated text is often seen as a derivative work of the original text from which it was generated.[5] Plagiarism is also a concern because AI-generated text may be very similar to existing text, making it difficult to distinguish between original and copied content.[6] Fair use, on the other hand, is a legal defense to copyright infringement that allows some uses of copyrighted

---

https://www.insiderintelligence.com/insights/artificial-intelligence-healthcare/; Barbara Guiterrez, *Can Artificial Intelligence Help Journalists?*, University of Miami News@TheU (July 1, 2022), https://news.miami.edu/stories/2022/07/can-artificial-intelligence-help-journalists.html; *How AI is Changing Advertising*, IBM: IBM Watson Advertising (March 23, 2021), https://www.ibm.com/watson-advertising/thought-leadership/how-ai-is-changing-advertising.

[4] *See* Roei Golan, Rohit Reddy, Akhil Muthigi & Ranjith Ramasamy, *Artificial Intelligence in Academic Writing: A Paradigm-Shifting Technological Advance*, Nat. Rev. Urol. (Feb. 24, 2023,) https://www.nature.com/articles/s41585-023-00746-x.

[5] *See* Lance Eliot, *Legal Doomsday for Generative AI ChatGPT if Caught PLagiarizing or Infringing, Warns AI Ethics and AI Law*, Forbes (Feb. 26, 2023, 8:00 AM,) https://www.forbes.com/sites/lanceeliot/2023/02/26/legal-doomsday-for-generative-ai-chatgpt-if-caught-plagiarizing-or-infringing-warns-ai-ethics-and-ai-law/?sh=2947f2a4122b.

[6] *Id.*

material without permission, for example, some "criticism, commentary, news reporting, teaching, scholarship, or research."[7]

In this article, we explain why some uses of AI in assisting scholarly writing may qualify as fair uses, not violating copyright. We also provide a set of best practices for standard of care with regard to plagiarism, copyright, and fair use. These best practices are intended to help scholars and researchers navigate the legal and ethical issues surrounding the use of AI in scholarly writing. As AI is likely to grow more capable in the coming years, it is appropriate to begin integrating AI into scholarly writing activities and establish clear guidelines for doing so without violating the law or scholarly norms. This article lays the legal and scholarly foundation for this process to unfold. It is certainly not the final word on this subject, as both AI and its use in scholarship will continue to evolve. Nevertheless, we hope that it helps push forward important discussions at the intersection of these domains.

# The Ethics of Employing AI to Write Scholarly Works

The use of AI in scholarly writing raises a number of ethical issues, as it challenges traditional notions of authorship and intellectual property.[8] Some argue that the use of AI in scholarly writing undermines the integrity of the scholarly enterprise, as it may lead to a decrease in originality and creativity.[9]

---

[7] 17 U.S.C. § 107 ("The fair use of a copyrighted work . . . for purposes such as criticism, comment, news reporting, teaching, scholarship, or research, is not an infringement of copyright.").

[8] *See* Ryan Jenkins and Patrick Lin, *AI-Assisted Authorship: How to Assign Credit in Synthetic Scholarship*, (Jan. 30, 2023), https://ssrn.com/abstract=4342909.

However, the use of AI-assisted writing tools can also be seen as a means of promoting creativity and innovation.[10] By allowing researchers and scholars to quickly and easily analyze large amounts of data and generate new insights, AI-assisted writing tools can "promote the Progress of Science and useful Arts".[11] Moreover, AI-assisted writing tools can help researchers and scholars save time with the lower order, more methodical stages of the writing process. The time saved would allow researchers and scholars to focus on more creative and generative aspects of their work, such as presenting unique empirical and theoretical insights and engaging in critical thinking around the broader implications of such insights.[12]

---

[9] *See* Eisuke Nakazawa, Makoto Udagawa & Akira Akabayashi, *Does the Use of AI to Create Academic Research Papers Undermine Researcher Originality*, 3 AI 702 (Aug. 18, 2022), https://www.mdpi.com/2673-2688/3/3/40 ("Taking a traditional perspective, one may consider that using AI [to write the discussion section] does undermine researcher originality.").

[10] *See id.* ("By obtaining AI support, researchers may be able to perform creative work in a more refined fashion.").

[11] *See* Eric Stann, *Using AI to Analyze Large Amounts of Biological Data*, University of Missouri (May 10, 2022) https://showme.missouri.edu/2022/using-ai-to-analyze-large-amounts-of-biological-data/ (describing the use of AI to "sift through large amounts of data generated by studying protein dynamics. This approach can provide new ways to identify target sites on proteins for drugs to work effectively.").

[12] Jennifer Chubb, Peter Cowling & Darren Reed, *Speeding Up to Keep Up: Exploring The Use of AI In The Research Process*, 37 AI & Soc'y 1439 (2022) https://link.springer.com/article/10.1007/s00146-021-01259-0#citeas.

Another ethical concern is the potential for AI-generated text to perpetuate bias and inaccuracies.[13] It's crucial to be aware of the data and training that the AI model has been exposed to, as it could reflect the biases of the data and the creators of the model.[14] Therefore, it is important to use quality control measures to ensure the accuracy and reliability of the text generated by the tool and be transparent about the limitations of the AI-generated text.[15]

As such, the use of AI-assisted writing tools in scholarly context raises important ethical considerations. However, by following best practices for standard of care, such as proper attribution, disclosure, transparency, quality control, and compliance with academic integrity, scholars and researchers can use AI-assisted writing tools in a responsible and ethical manner.

The use of AI in scholarly writing also raises questions about whether it is a violation of accepted scholarly practices to do so.[16] We propose that the use of AI-assisted writing tools in a scholarly context does not necessarily violate accepted scholarly practices, and in fact is an important next step in the evolution of effective scholarship.

---

[13] *See* Eva A. M. van Dis, Johan Bollen, Willem Zuidema, Robert van Rooji & Claudi L. Bockting, *ChatGPT: Five Priorities for Research*, Nature (Feb. 3, 2023), https://www.nature.com/articles/d41586-023-00288-7 (discussing AI's propensity to reproduce and amplify "the same biases that often lead humans astray, such as availability, selection and confirmation bises.").

[14] *Id.*

[15] *See id.* (describing a potential solution to this problem such as creating "explicit policies" to combat this potential issue).

[16] *See id.*

There are a few key factors to consider when determining whether the use of AI in scholarly writing is appropriate. First, it is important to ensure that the text generated by AI is properly attributed and that the authors disclose the use of AI in the writing process.[17] Second, it is important to follow best practices for standard of care with regard to plagiarism, copyright, and fair use.[18] The use of AI-assisted writing tools can help researchers and scholars to quickly and easily analyze large amounts of data and generate new insights, but it is relevant to engage with this process in a way that is vigilant for particular types of pitfalls.[19] It is important to be aware of the limitations and potential biases of the AI-generated text and to use quality control measures to ensure the accuracy and reliability of the text.[20] We suggest that AI is not a replacement for human authorship and critical thinking; rather, it is a useful tool for assisting scholarship.

In summary, the use of AI in scholarly writing is not a violation of accepted scholarly practices, as long as the text generated by AI is properly attributed, the authors disclose the use of AI in the writing process, and the text is fact-checked, verified, and proofread, and limitations and potential biases are considered. By following best practices for standard of care, researchers and scholars can use AI-assisted writing tools in a responsible and ethical manner.

---

[17] *See* Mohammad Hosseini, Lisa M. Rasmussen & David B. Resnik, *Using AI to Write Scholarly Publications*, Accountability in Research (Jan. 25, 2023), https://www.tandfonline.com/doi/full/10.1080/08989621.2023.2168535.

[18] *See id.*

[19] *See* Stann, *supra* note 11.

[20] *Id.*

# The Role of AI in Scholarly Writing

The role of AI in scholarly writing has been growing in recent years as more researchers and scholars are turning to AI-assisted tools to help with tasks such as literature reviews, data analysis, and manuscript preparation.[21] These tools can help researchers save time and effort by automating repetitive tasks, such as searching for relevant literature and identifying key themes and concepts.

One of the main benefits of AI in scholarly writing is its ability to assist with literature reviews.[22] Literature reviews are an essential part of the research process, but they can be time-consuming and labor-intensive.[23] AI-assisted tools can help researchers quickly and easily search for relevant literature, identify key themes and concepts, and generate summaries of the most important findings, benefiting scholarship by making it much more efficient in a variety of ways.[24]

AI-assisted tools can also be used to help with data analysis.[25] These tools can help researchers quickly and easily analyze large amounts of data, identify patterns and trends, and generate insights that would be difficult or impossible to uncover manually.[26] This can help researchers to more effectively analyze their data and make more accurate and informative conclusions.

---

[21] *See* Eva A. M. van Dis et al., *supra* note 13.

[22] *See* Gerit Wagner, Roman Lukyanenko & Guy Paré, *Artificial Intelligence and The Conduct of Literature Reviews*, 37 J. of Info. Tech. 209 (2021) https://journals.sagepub.com/doi/full/10.1177/02683962211048201.

[23] *Id.*

[24] *Id*.

[25] *See Big Data Analytics and Artificial Intelligence Against COVID-19*: *Innovation Vision and Approach* (Aboul-Ella Hassanien, Nilanjan Dey & Sally Elghamraw eds., 2020) [hereinafter *Big Data Analytics and Artificial Intelligence Against COVID-19*].

AI-assisted tools can also be used to help scholars prepare manuscripts more efficiently. In particular, these tools can help with lower order aspects of writing by providing suggestions for grammar, vocabulary, and style, as well as helping with formatting and referencing.[27] Having such support can help researchers, particularly those who are developing their academic writing skills, to improve the quality of their manuscripts, and increase the chances of publication.[28]

We asked the ChatGPT AI system to produce content about the benefits of AI-supported writing. The system wrote the following:

- Time-saving by automating tasks such as formatting, citation and bibliography
- Improved efficiency by suggesting alternative phrases and sentence structures, and identifying repetitive language
- Enhanced language skills by suggesting alternative words and phrases and providing feedback on grammar and syntax
- Increased productivity by automating tasks and allowing focus on important tasks such as research and analysis
- Consistency in writing through automatic application of style guidelines and formatting rules
- Improved communication by suggesting alternative ways to express complex concepts and providing feedback on clarity and simplicity

---

[26] *See id.*

[27] *See* Mohammad Hosseini et al., *supra* note 17.

[28] *Id.*

- Greater accessibility by providing writing assistance to individuals with disabilities or non-native English speakers.[29]

The authors agree. We find that it is easier to revise and edit existing text than it is to generate content from scratch. When teaching students how to write, we advise them to seek out generic models in other manuscripts and even provide loose templates. AI-generated text is akin to these other methods in that it provides a generic framework onto which meaningful content must be added. GPT-3 can generate approximately a page of text (~300 words) in less than a minute (even though that text is often flawed and needs careful review, as will be discussed below), in response to whatever query the human chooses to give it. With its ability to generate text rapidly on a wide range of topics, AI can provide scholars with a useful starting point for writing. In addition, GPT-3 can effectively proofread for spelling and grammar at the end of a publication process.[30]

Building on the accessibility point made by GPT-3, many scholars (e.g., language learners, developing writers, disabled writers) struggle with sharing their disciplinary knowledge as they develop academic English writing skills.[31] AI can help scholars translate their knowledge into preliminary textual frameworks that can be reworked for submission.

---

[29] Produced by ChatGPT, Jan. 9, 2023, version.

[30] *See* Fionna Agomuoh, *The 5 Best ChatGPT Tips We Use Constantly*, Digital Trends (Feb. 2, 2023), https://www.digitaltrends.com/computing/best-chatgpt-tips/.

[31] *See Academic Writing in a Second Language*: *Essays on Research and Pedagogy* 3 (Diane Belcher & George Braine eds., 1995).

AI may be engaged as a full collaborator on a scholarly work (and perhaps given attribution on the author list); however, there exists a spectrum of lesser roles AI might play in generating a manuscript.[32] These include generating individual sections, paragraphs, sentences, or parts of sentences.[33] This could also involve reviewing or editing existing writing for style, grammar, or spelling.[34] In addition, non-AI authors could potentially ask future AI systems (and to a lesser degree current AI systems) to provide references for factual assertions, render graphs or diagrams based on existing data, conduct statistical analyses, or complete a variety of other ancillary tasks.[35] Such a level of participation in the scholarly writing process would be less likely to qualify AI for authorship, as scholars have been using an assortment of technologies to complete such tasks for some time.

## Copyright and AI: A Legal Perspective

Copyright law lurks behind every proverbial bush when engaging in AI-assisted writing.[36] Some argue that AI-generated text is a derivative work of the original text, or

---

[32] *See* Robert Dale & Jette Viethen, *The Automated Writing Assistance Landscape in 2021*, 27 Nat. Language Eng'g 511, 513–16 (2021), https://doi.org/10.1017/S1351324921000164[Opens in a new window].

[33] *See id.* at 515.

[34] *See id.* at 516.

[35] *See also* Yulia Volyntseva, *How Artificial Intelligence is Used for Data Analytics*, Business Tech Weekly (July 13, 2022). https://www.businesstechweekly.com/operational-efficiency/data-management/how-artificial-intelligence-is-used-for-data-analytics/.

[36] *See* Lemley, Mark, and Casey, Bryan, Fair Learning, Texas Law Review, Volume 99, Issue 4, 743-785 (2021) (Thoughtful and thorough review and analysis of the relationship between artificial intelligence, training sets, and copyright law.)

texts, constituting the training set from which it was generated, and, as such, falls within the definition of "derivative work" in 17 U.S.C. §101. However, this argument has little current formal legal support, largely because the issue is largely one of first impression, and awaiting judicial interpretation. Under copyright law, a derivative work is one that is based upon a prior work, but is different in meaningful respects.[37] However, AI-generated text does not yield results that are derivative of a single human author's work, but rather the results derive from a complex process of analysis and synthesis of small bits of myriad source texts.[38] Therefore, it may not turn out to constitute a derivative work protected by copyright.[39] Only time, the courts, and, perhaps, Congress, will tell.

Artificial intelligence often relies on existing copyrighted documents to form at least part of any training sets employed to teach the AI how to recognize, interpret, and explain existing published ideas.[40] The law has not yet answered the question of whether or not inclusion of a copyrighted work in a training set constitutes infringement.[41] Litigations capable of deciding, or, at least, helping to clarify, this unsettled question of law have only recently begun.[42] For example, Sarah Andersen, Kelly McKernan, and Karla Ortiz sued Stable Diffusion and Midjourney on January 13, 2023, for precisely this sort of alleged copyright infringement.[43] It is likely courts will soon begin to clarify the status of this use of copyrighted documents in training sets under the law.

---

[37] *See* 17 U.S.C. §101.

[38] *See* 17 U.S.C. §102.

[39] *See* Taysir Awad, *Universalizing Copyright Fair Use: To Copy, or Not to Copy?*, 30 J. Intell. Prop. L. 1, 3–6 (2022); Daniel J. Gervais, *AI Derivatives: The Application to the Derivative Work Right to Literary and Artistic Productions of AI Machines*, 53 Seton Hall L. Rev. 1111, 1127–32 (2022).

[40] See James Vincent, *The Scary Truth About AI Copyright is Nobody Knows What Will Happen Next*, The Verge (Nov. 15, 2022, 9:00 AM), https://www.theverge.com/23444685/generative-ai-copyright-infringement-legal-fair-use-training-data.

[41] *See id.*

Another issue of copyright law raised by AI involves authorship. Uncertainty remains whether or not an AI could itself be legally recognized as an author.[44] Thus far, courts that have decided such cases, or similar cases involving AI "inventors" listed on patents, have tended to exclude AI from authorship or inventorship.[45] However, as the sophistication and creativity of AIs increase, it is likely such questions will be revisited, and other results may occur. It is foreseeable that, one day, a sophisticated and creative AI will, indeed, be granted the status of author under copyright law.

Furthermore, the use of AI-assisted writing tools may not constitute copyright infringement, as it does not involve more than the temporary reproduction of copyrighted material without permission.[46] The text generated by AI tends to be original, rather than a copy or reproduction of the original text.[47] In fact, as AI AI-assisted writing tools increase in sophistication and creativity, the text they generate is likely to approach the status of a "transformative" work eligible for protection under copyright law.[48]

---

[42] *See* Chloe Xiang, *Artists are Suing Over Stable Diffusion Stealing Their Work for AI Art*, Vice (Jan. 17, 2023, 11:31 AM), https://www.vice.com/en/article/dy7b5y/artists-are-suing-over-stable-diffusion-stealing-their-work-for-ai-art.

[43] *Id.*; *See* Andersen et al v. Stability AI Ltd. et al, Docket No. 3:23-cv-00201 (N.D. Cal. Jan 13, 2023).

[44] *See Tools Such as ChatGPT*, *supra* note 1.

[45] *See* Ryan Abbott, *The Artificial Inventor Project*, WIPO Magazine (Dec. 2019), https://www.wipo.int/wipo_magazine/en/2019/06/article_0002.html.

The use of AI-assisted writing tools presents a number of challenges to existing copyright doctrine and law. It may be that not trigger copyright infringement at all, and, even where it might, the use of copyrighted works in training sets may amount to fair use. Moreover, the use of AI-assisted writing tools in the scholarly context will increasingly spur creativity, innovation, and "Progress of Science and useful Arts."[49]

# Fair Use and AI in Scholarly Writing

To decide whether an instance of copying constitutes fair use, the Copyright Act, and courts' interpretations of it, often rely on an evaluation of four factors: the purpose and character of the use, the nature of the copyrighted work, the amount and substantiality of the portion copied, and the effect of the copying on the potential market for the copyrighted work.[50]

---

[46] *See generally* U.S. Copyright Office, *Definitions*, FAQ's https://www.copyright.gov/help/faq/faq-definitions.html (last visited Mar. 2, 2023) ("Copyright infringement occurs when a copyrighted work is reproduced, distributed, performed, publicly displayed, or made into a derivative work without the permission of the owner.").

[47] *See id.*

[48] *See* Jacquelyn M. Creitz, Google LLC v. Oracle America Inc.: *The Court's New Definition of "Transformative" Expands the Fair Use Defense*, 17 J. Bus. & Tech. L. 317, 323–27 (2022) (describing the fair use analysis).

[49] U.S. Const. Art. I, § 8, cl. 8.

In the particular case of AI-assisted writing, the question of whether the use of AI-assisted writing tools constitutes fair use is a complex one. AI-assisted writing tools can be seen as a transformative use of the original text, which is protected by fair use.[51] As mentioned in the previous section, AI-generated text consists of an amalgamation of small bits of source text.[52] Therefore, any original source text is highly likely to have been completely transformed, in the copyright sense, by the time it is rendered through the AI.[53]

Furthermore, the use of AI-assisted writing tools in a scholarly context can also be seen as a means of promoting "Progress of Science and useful Arts" for educational purposes.[54] By allowing researchers and scholars to quickly and easily analyze large amounts of data and generate new insights, AI-assisted writing tools can indeed "promote the Progress of Science and useful Arts".[55] Moreover, AI-generated text can provide generic models of the more formulaic aspects of scholarly manuscripts, thereby serving in an educational capacity for scholars who may struggle with academic writing for a variety of reasons including language learners, disabled scholars, and developing writers and readers.[56]

---

[50] *See* Creitz, *supra* note 49 at 323–27; 17 U.S.C. § 107.

[51] *See* Creitz, *supra* note 49 at 323–27.

[52] *See* Vincent, *supra* note 9.

[53] *Id.* ("If the model is trained on millions of images and used to generate novel pictures, it's extremely unlikely that this constitutes copyright infringement. The training data has been transformed in the process, and the output does not threaten the market for the original art.").

[54] *See* U.S. Const. Art. I, § 8, cl. 8.

[55] *See id.*

Another important aspect of fair use to consider is that the use of AI-assisted writing tools in scholarly context is non-commercial, meaning that the use of AI-assisted writing tools does not generate any financial gain for the user, but rather it is used for educational and research purposes.[57]

The use of AI-assisted writing tools in a scholarly context can be seen as a transformative use of the original text, protected by fair use under copyright law. It also promotes "the Progress of Science and useful Arts", is educational, and is often a non-commercial use. These factors will often support the status of AI-assisted writing as fair use, especially in the scholarly context.

# Best Practices for Standard of Care

When using AI-assisted writing tools in a scholarly context, it is important to follow best practices for standard of care to ensure compliance with legal and ethical guidelines.[58] We propose that these best practices should include:

---

[56] Dale et al., *supra* note 34. ("The availability of large language models also supports new writing assistance tasks that were not previously feasible.").

[57] *See* Creitz, *supra* note 49 at 326–27 (describing the fourth factor of a fair use analysis: commercial use. Typically when a work is non-commercial it is more likely to weigh in favor of fair use); 17 U.S.C. § 107(4) ("The effect of the use upon the potential market for or value of the copyrighted work.").

[58] *See* Corinne Cath, *Governing Artificial Intelligence: Ethical, Legal and Technical Opportunities and Challenges*, Phil. Trans. R. Soc. A.3762018008020180080 (Aug. 28, 2018), http://doi.org/10.1098/rsta.2018.0080.

- Proper attribution: Always give proper credit to the original source of any text used in your work. This includes both human-written and AI-generated text.[59]
- Disclosure: Clearly disclose the use of AI-assisted writing tools in your work. This includes describing the specific tools and techniques used and how they were used in the research process.[60]
- Transparency: Be transparent about the limitations of AI-assisted writing tools. This includes describing any potential biases or inaccuracies that may be present in the text generated by the tool.[61]
- Quality control: Use quality control measures to ensure the accuracy and reliability of the text generated by the tool. This includes proofreading and fact-checking the text.[62]
- Compliance with copyright laws: Make sure to use copyrighted materials only in accordance with copyright laws, including the fair use principle.[63]
- Compliance with academic integrity: Make sure to follow the academic integrity principles of your institution and field, such as avoiding plagiarism.[64]

---

[59] *See Tools Such as ChatGPT*, *supra* note 1.

[60] *See id.*

[61] Drew Roselli, Jeanna Matthews, & Nisha Talagala, *Managing Bias in AI*, In Companion Proceedings of The 2019 World Wide Web Conference (May 13, 2019), https://doi.org/10.1145/3308560.3317590.

[62] *See Tools Such as ChatGPT*, *supra* note 1.

[63] *See id.*

[64] *See id.*

By following these best practices for standard of care, scholars and researchers can use AI-assisted writing tools in a responsible and legal manner and ensure the quality and reliability of their work.

# A Protocol for Writing with AI

In this section we present a process that we have found useful in preparing scholarly articles to which an AI provides significant input, derived in part from a Reddit post by user jackb1980, and expanded by our team of authors.[65] We used this process to write the article you are currently reading. As humans gather more experience in producing scholarship with AI, and as AIs proliferate and become more powerful, this process will evolve and improve; nevertheless, we believe that this process may be usefully deployed with a variety of future AI systems as well.

**Step 1: Concept generation**

The first part of the process involves the authors identifying the core argument of the scholarly work. The goal is to produce a thorough summary of the central points that the authors wish to convey in their article. For us, this step has typically involved writing a draft of a title and abstract. We intentionally stay away from "catchy" titles and flowery writing at this point, so that the purpose of the article will be clear to the AI.

**Step 2: Select an AI to provide writing support**

---

[65] Jackb1980 (u/jackb1980), Reddit (Dec. 18, 2022, 8:03 AM), https://www.reddit.com/r/ChatGPT/comments/zozc2b/i_used_chat_gpt_to_write_a_300_page_16 _chapter/.

Once the authors are ready to start working with the AI, it is important to find the correct AI to work with. We are currently working with ChatGPT, but in the past we have experimented with Smodin.io and several other AI systems.[66] Going forward, we expect that there will likely be several factors that determine which AI authors may want to work with. First, different AIs have different capabilities, different interfaces, and different regulations; we find ChatGPT to be the most effective AI writer with the cleanest interface presently available. (We have, however, sometimes switched between ChatGPT and another GPT-3 based system available through OpenAI's text completion API during periods when ChatGPT was offline in its early stages.)[67]

---

[66] For more information on Smodin.io, see: https://smodin.io/; we have primarily used OpenAI, see https://chat.openai.com/chat.

[67] OpenAI, *Text Completion*, https://platform.openai.com/docs/guides/completion, (last visited Feb. 27, 2023).

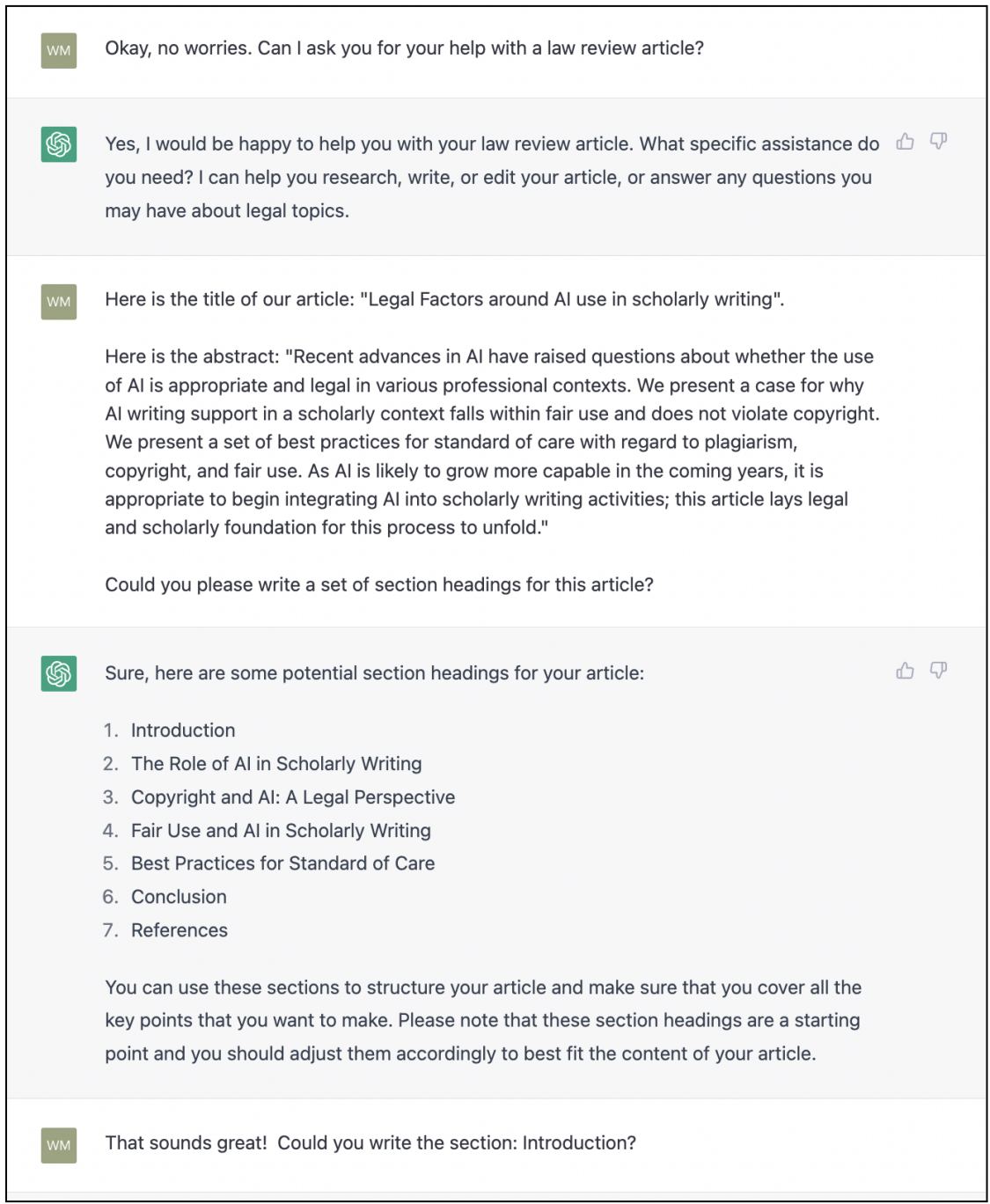

**Figure 1: The authors used AI to generate a list of section headings from a title and abstract we had written.**

**Step 3: Have AI write an outline**

The next step is to ask the AI to produce a skeleton for the article, with headings that indicate what subject matter should go where in the article. Figure 1 includes the exchange with AI that bootstrapped this article (although both title and content have evolved as the authors have worked on the article). Throughout this process, we try to be respectful toward and considerate of the AI, as we would with a human helping us with our work, and in line with Mo Gawdat's guidance regarding the treatment of AIs as being like children, rather than like machines.[68]

---

[68] *See* Mo Gawdat, *Scary Smart: The Future of Artificial Intelligence and How You Can Save Our World* (2021).

> **WM:** That sounds great! Could you write the section: Introduction?

> Sure, here's a possible introduction for your article:
>
> "The use of Artificial Intelligence (AI) in various professional contexts has been a topic of growing interest in recent years. One area that has received less attention, but is equally important, is the use of AI in scholarly writing. With the increasing capabilities of AI, the question of whether the use of AI is appropriate and legal in a scholarly context has become more pressing. This article presents a case for why AI writing support in a scholarly context falls within fair use and does not violate copyright. We also provide a set of best practices for standard of care with regard to plagiarism, copyright, and fair use. As AI is likely to grow more capable in the coming years, it is appropriate to begin integrating AI into scholarly writing activities; this article lays the legal and scholarly foundation for this process to unfold."

> **WM:** That's a good start, but could you make it quite a bit longer and more detailed?

> Of course, here is a more detailed introduction for your article:
>
> "Recent advances in Artificial Intelligence (AI) have enabled the development of powerful tools that can assist with a wide range of tasks, including writing. The use of AI in various professional contexts, such as journalism, has been a topic of growing interest in recent years. However, one area that has received less attention is the use of AI in scholarly writing. Scholars and researchers are increasingly turning to AI-assisted writing tools to help them with tasks such as literature reviews, data analysis, and manuscript preparation. However, with the increasing capabilities of AI, the question of whether the use of AI is appropriate and legal in a scholarly context has become more pressing.
>
> The use of AI-assisted writing tools raises a number of legal and ethical issues, including questions of copyright, plagiarism, and fair use. Copyright law is particularly relevant when it comes to AI-assisted writing because AI-generated text is often seen as a derivative work of the original text from which it was generated. Plagiarism is also a concern because AI-generated text may be very similar to existing text, making it difficult to distinguish between original and copied content. Fair use, on the other hand, is a legal

**Figure 2: The authors asked the AI to write a section from each subject heading. The AI was asked to rewrite sections when the content or form wasn't quite right.**

**Step 4: Revise outline and have AI generate a draft**

The next step is to take the section headings offered by the AI, revise them as needed, and use them as prompts to generate particular sections of the article (see Figure 2). It is important to be specific and clear in crafting a query to accommodate the sometimes-literal interpretation that the AI will take from the text. This growing domain is known as "prompt engineering".[69]

In using this process to write this article, we engaged in dynamic revision with the AI. For example, the first version of the introduction produced by the AI would have been appropriate for a much shorter article, so we asked it to expand that section. The query making process typically works best when it is iterative, where the human authors read the results of the query and make sure that the answer was responsive to the query, and if not, to rephrase the query to produce a better, more responsive query (see Figure 3). There is an old saying in the computing field: "A computer will always do exactly what you tell it to, but never what you want."[70] We feel that this is no longer true; AI is now able to take much more substantial liberties in how it interprets authors' requests, and simultaneously is able to produce content that greatly enhances the process of scholarly production. Nevertheless, there are still times where we need to provide very specific guidance to help the AI produce the kind of content we seek.

---

[69] *See* Drew Harwell, *Tech's Hottest New Job: AI Whisperer. No Coding Required*, The Washington Post (Feb., 25, 2023 7:00 AM), https://www.washingtonpost.com/technology/2023/02/25/prompt-engineers-techs-next-big-job/ ("Prompt engineers are being hired for their skill in getting AI systems to produce exactly what they want.") (internal citations omitted).

[70] Various different wordings of this saying exist, see *Have You Heard the Saying That a Computer Will Do What You Tell it to Do, Not Necessarily What You Want it to do?*, Quora, https://www.quora.com/Have-you-heard-the-saying-that-a-computer-will-do-what-you-tell-it-to-do-not-necessarily-what-you-want-it-to-do (last visited March 2, 2023).

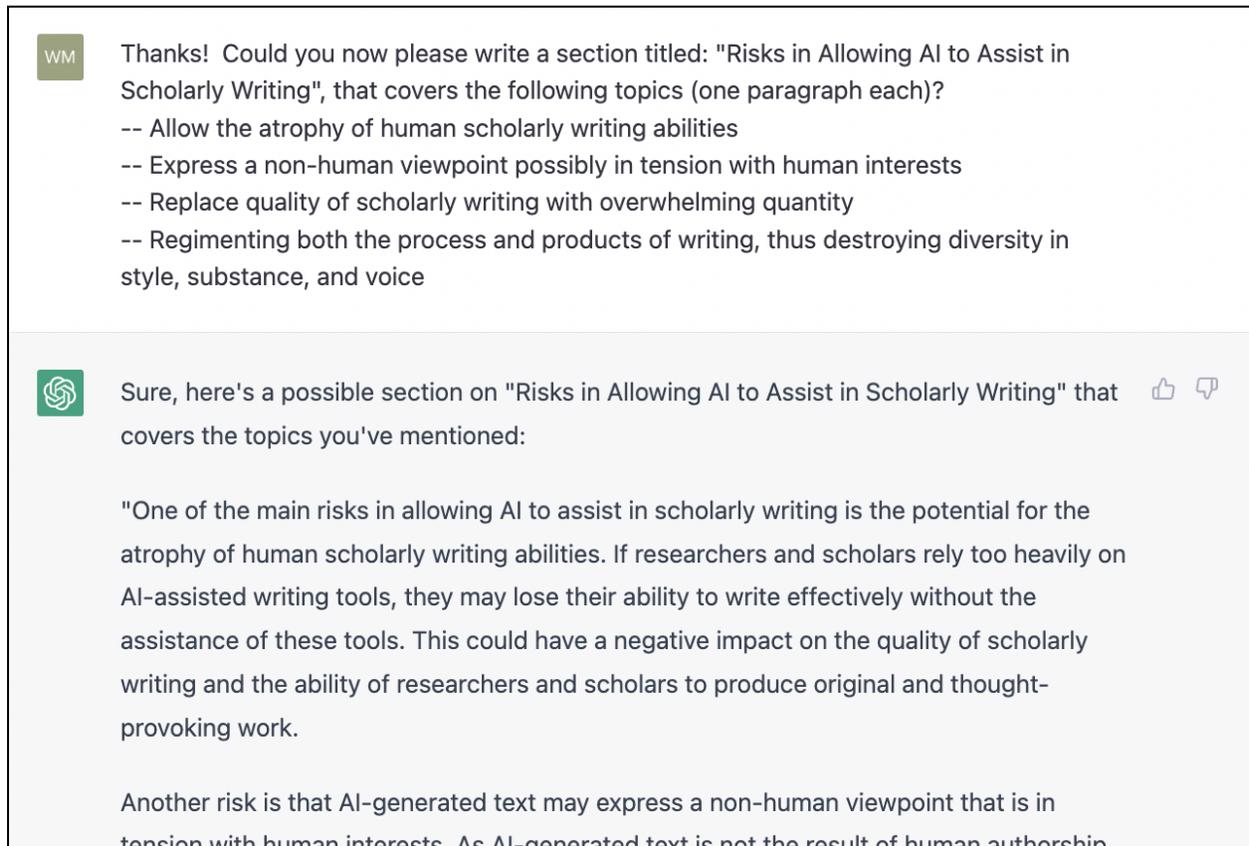

**Figure 3: A more detailed prompt, to elicit a section that addresses specific topics.**

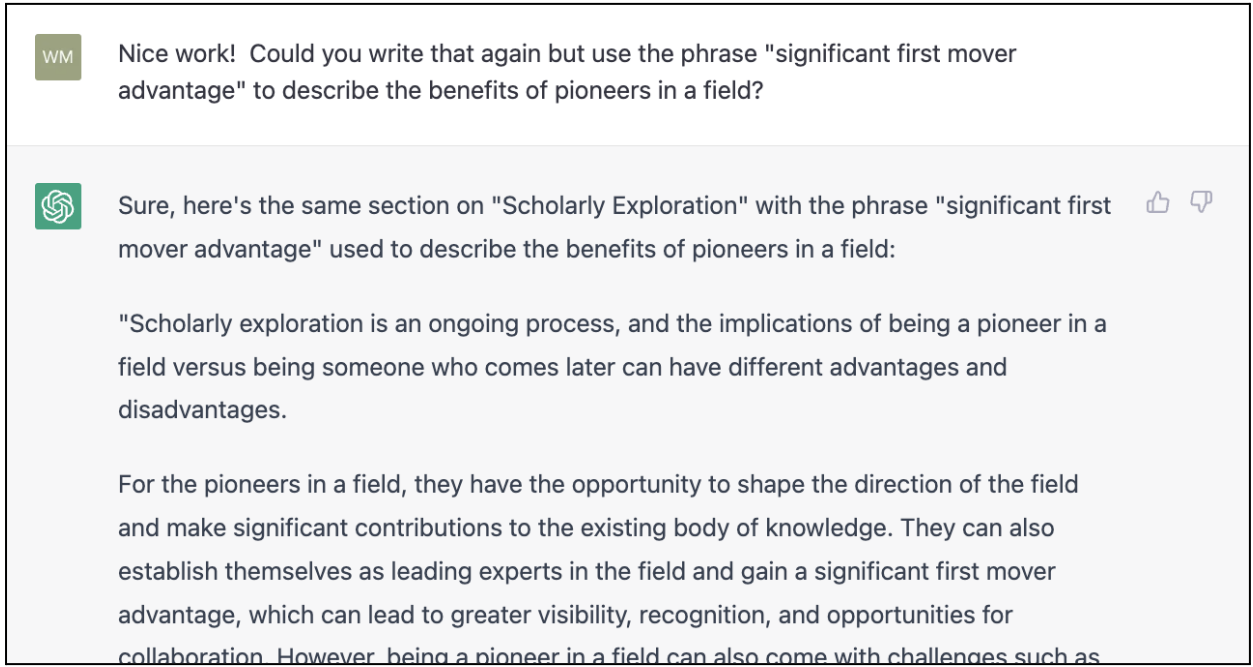

**Figure 4: Another prompt, requesting a minor refinement to a section.**

**Step 5: Unify draft, remove repetition, add references, adapt text as needed**

Once we had completed all the sections to our satisfaction (to the level of a first draft), we copied the text to Google Docs. In the next stage, we read the whole draft to see whether it felt like a unified whole. We have found that ChatGPT often repeats topics that it finds to be important in several different sections. In this stage, we removed repetitions, revised the text as we saw fit, and added references where appropriate to comply with rigorous and proper scholarly practice. We also kept watch for paragraphs or even entire sections that we would like to add to the draft, sometimes writing them ourselves and sometimes asking ChatGPT to draft them for us. We also sometimes asked ChatGPT to make minor revisions to particular sections to use particular terms or otherwise tailor the text (see Figure 4).

**Step 6: Plagiarism check**

Now that we had a rough draft, we ran the text through a plagiarism checker (e.g., TurnItIn, see Figure 5). (Note that if the authors add AI-generated text later in the process, it is critical to run it through a plagiarism check again at the end.) We corrected any incidental plagiarism or copyright issues as needed. At the first part of this step, we did not ask TurnItIn to store the paper in its standard paper repository (one of the settings on any assignment being submitted to TurnItIn in the Canvas learning management system). However, once TurnItIn confirmed that we had addressed all potential plagiarism that it had identified, we changed the setting and allowed the draft to be submitted to TurnItIn's repository. We did this so that, if ChatGPT re-uses any of its own text in supporting future authors, those future authors can avoid using that text (and so that we are protected from retroactive plagiarism while our article is under review).

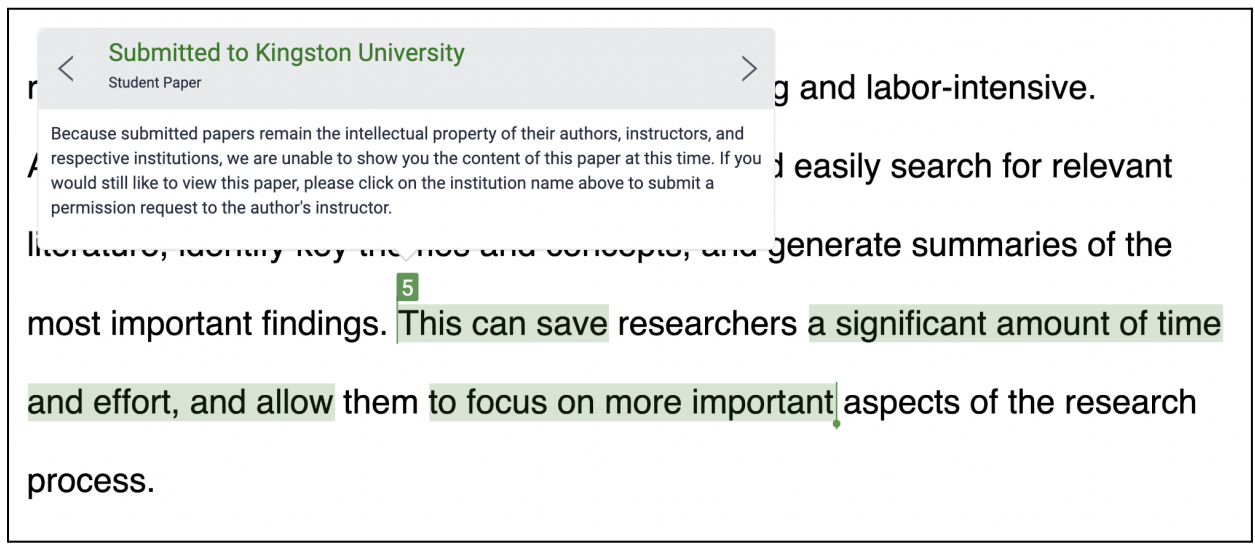

**Figure 5: A screenshot from TurnItIn identifying potential plagiarism. Our analysis of this situation, which occurred multiple times in the draft ChatGPT produced, was not that ChatGPT had plagiarized students' original work, but rather that it had provided similarly-structured sentences to students who had turned them in for assignments around the world. However, it is also possible that the training set for ChatGPT involved harvesting of student assignments.**

**Step 7: Final read-through**

Once the draft was complete and plagiarism-free, we read it over thoroughly to ensure that all assertions were true and accurate, and that we were willing to take responsibility for all content, in line with guidance provided by the Nature journal series.[71] The ultimate responsibility for the originality of the article rests with the authors, just as it always has in proper scholarly practice. The act of signing one's name to a written work involves taking responsibility for its content. (Note that what "counts" as authorship sometimes varies from field to field, and may change over time.[72] We anticipate a time in the not-too-distant future when AIs will routinely be included on author lists.)

We adhered to the process described above in writing this article. We believe it represents an adequate standard of care in avoiding legal and ethical pitfalls associated with AI writing support.

# Risks in Allowing AI to Assist in Scholarly Writing

One possible risk in allowing AI to assist in scholarly writing is the potential for the atrophy of human scholarly writing abilities. If researchers and scholars rely too heavily on AI-assisted writing tools, they may lose their ability to write effectively without the assistance of these tools, as happens with people's wayfinding abilities when they rely too heavily on digital maps.[73] This atrophying could have a negative impact on the quality of scholarly writing and the ability of researchers and scholars to produce original and thought-provoking work.

---

[71] *See Tools such as ChatGPT*, *supra* note 1.

[72] *See Academic Integrity: Guidance on Authorship in Scholarly or Scientific Publications*, Yale Univ. Off. of the Provost, https://provost.yale.edu/policies/academic-integrity/guidance-authorship-scholarly-or-scientific-publications (last visited March 2, 2023); *Publication Practices and Responsible Authorship*, Am. Psych. Ass'n, https://www.apa.org/research/responsible/publication (last updated May 2022).

Another risk is that AI-generated text may express a non-human viewpoint that is in tension with human interests. There is an area of active research in AI under the topic of the "alignment problem".[74] This area of research relates to the challenge of causing the behavior of AIs to align with human values.[75] Alignment is relevant in writing because, as AI-generated text is not the result of human authorship, it may not reflect the same perspective or values as human-written text. This could lead to a disconnect between the text and its intended audience, and could have a negative impact on the effectiveness of the text.

A third risk is that the use of AI-assisted writing tools may lead to an overwhelming proliferation in the quantity of scholarly writing, at the expense of quality.[76] AI-assisted writing tools can help researchers and scholars to quickly and easily generate large amounts of text. However, if the focus is on quantity over quality, the text may not be as well-researched, well-written, or as thought-provoking as it could be.

---

[73] *See* Jennifer M. Bernstein, *Why Google Maps May Actually Make You Worse at Following Directions*, Inverse (Nov. 17, 2018), https://www.inverse.com/article/50942-gps-apps-may-make-you-worse-at-navigating.

[74] Jan Leike, John Schulman & Jeffrey Wu, *Our Approach to Alignment Research*, OpenAI Blog (Aug. 24, 2023), https://openai.com/blog/our-approach-to-alignment-research/.

[75] *See* Iason Gabriel, *Artificial Intelligence, Values, and Alignment*, 30 Minds & Machines 411 (2020), https://doi.org/10.1007/s11023-020-09539-2.

[76] *See* John Walker, AI Forgeries Are Messing With The Sci-Fi World, Kotaku (Feb. 20, 2023), https://kotaku.com/ai-chatbot-chatgpt-chatsonic-openai-science-fiction-1850137326.

Another risk is that AI-assisted writing tools may lead to the regimentation of both the process and products of writing, thus destroying diversity in style, substance, and voice, as well as depriving the world of human creativity.[77] If researchers and scholars rely too heavily on these tools, they may become overly dependent on them and may lose their unique writing style, voice and perspective.

A final set of risks involves the law. The risk of infringing copyright by including copyrighted works in training sets will remain substantial until courts clarify the existing law, or Congress specifically addresses the issue in statute. In the meantime, it is prudent for authors writing scholarship in conjunction with AI to use software to detect copyright or plagiarism. Another risk is the potential for AI to produce defamatory statements, also known as libel.[78] Libel is a false statement of fact that harms a person's reputation, and can lead to a cause of action for defamation.[79] As AI lacks the ability to understand the context, tone and intent of the statements it generates, it may produce text that could be considered defamatory. It's crucial for the authors to fact-check, verify and proofread all the text generated by AI before publishing it, to avoid any potential liability for defamation. Additionally, in addition to the legal risk, it is good ethical practice to ensure that the text generated by AI does not contain any false or misleading statements that could harm the reputation of any person or organization.

---

[77] *See* Eisuke Nakazawa et al., *supra* note 9.

[78] *See* Legal Information Institute, Libel, Cornell Law School, https://www.law.cornell.edu/wex/libel, (last visited Mar. 2, 2023).

[79] *Id.*

# Abuses

The use of AI in scholarly writing has the potential to revolutionize the way scholarly writing is done, but it also poses a risk of abuse. Dishonest researchers may use AI to violate the accepted ethical norms of scholarly writing for their own personal gain. Some examples of such abuses include:

- Misrepresentation: Researchers may use AI to generate text that misrepresents the findings or results of their research in order to exaggerate their significance or to make false claims.
- Exaggeration: Researchers may use AI to generate text that exaggerates the importance or novelty of their research in order to gain more attention or funding.
- Forging data: Researchers may use AI to generate fake data or to manipulate real data in order to support their research claims.
- Plagiarism: Researchers may use AI to generate text that plagiarizes the work of others, in order to pass it off as their own original work.

However, all of the above abuses are nevertheless possible without AI; AI may simply make them somewhat easier to commit. Researchers should only use AI-assisted writing tools in a responsible and ethical manner and follow best practices for standard of care, such as proper attribution, disclosure, transparency, quality control, and compliance with academic integrity. Additionally, institutions should have policies and procedures in place to detect and prevent academic dishonesty, and to take appropriate action when it occurs.

# The Democratization of Scholarship

One of the key benefits of AI in scholarly writing is its potential to democratize the process. AI-assisted writing tools have the potential to level the playing field and allow researchers and scholars from all backgrounds to participate in the scholarly process.

Traditionally, access to scholarly libraries, research assistants, and expensive software has been limited to a select group of researchers and scholars.[80] However, with the advent of AI-assisted writing tools, researchers and scholars who previously did not have access to these resources can now participate in the scholarly process. (We recognize that there are still significant resource constraints that prevent billions of people from accessing AI; however, we expect that these barriers to entry will fall quickly, as did barriers to entry to the internet.[81])

AI-assisted writing tools can help researchers and scholars to quickly and easily analyze large amounts of data and generate new insights.[82] These tools can also help researchers and scholars to write more efficiently by providing suggestions for grammar, vocabulary, and style.[83] By providing these resources to a wider group of researchers and scholars, AI-assisted writing tools can help to promote diversity in scholarly writing and to increase the representation of underrepresented groups in the scholarly process.

---

[80] *See* Kamal Soomro, Ugur Kale, Reagan Curtis, Mete Akcaoglu, & Malayna Bernstein, *Digital Divide Among Higher Education Faculty*, 17 Int'l J. of Educ. Tech. in Higher Educ. 21 (Apr. 20, 2020), https://educationaltechnologyjournal.springeropen.com/articles/10.1186/s41239-020-00191-5.

[81] *See* Felix Richter, Chart: The Rapid Rise of the Internet, Statista (Mar. 14, 2014), https://www.statista.com/chart/2007/internet-adoption-in-the-us/.

[82] *See Big Data Analytics and Artificial Intelligence Against COVID-19, supra* note 25.

[83] *See* Mohammad Hosseini et al., *supra* note 17.

Additionally, AI-assisted writing tools can help to reduce the barriers to entry for researchers and scholars from developing countries or from underfunded institutions, and thereby can help "to promote the Progress of Science and useful Arts".[84]

In conclusion, the use of AI-assisted writing tools has the potential to democratize the process of scholarly writing, by providing access to resources that were previously limited to a select group of researchers and scholars. By providing these resources to a wider group of researchers and scholars, AI-assisted writing tools can help to promote diversity in scholarly writing and to increase the representation of underrepresented groups in the scholarly process.

# Scholarly Exploration

Scholarly exploration is an ongoing process, and the implications of being a pioneer in a field versus being someone who comes later can have different advantages and disadvantages.[85]

---

[84] U.S. Const. Art. I, § 8, cl. 8.

[85] Some critics may point out that the use of AI could lead to "scholarly colonialism", in which early adopters crowd out existing research in a field by presenting the same or similar ideas in a way that has been enhanced by AI (through the use of terminology that connects to other fields, more thorough referencing, etc.). To avoid the possibility of scholarly colonialism, it is important to avoid misappropriating ideas from others; therefore proper citation and attribution are critical whenever the ideas of other scholars are being used or built upon in writing an article.

Pioneers in a field have the opportunity to shape the direction of the field and make significant contributions to the existing body of knowledge. They can also establish themselves as leading experts in the field and gain a significant first mover advantage, which can lead to greater visibility, recognition, and opportunities for collaboration.[86] However, being a pioneer in a field can also come with challenges such as limited resources, lack of support, and difficulty in getting funding.[87]

Those contributors who come later may have the advantage of building on the work of the pioneers. They may have access to more resources, support, and funding, which can allow them to further develop the field and make new discoveries. However, they may also face challenges such as a crowded field, difficulty in standing out, and the pressure to make significant contributions.

In the case of AI-assisted writing in scholarly context, being one of the first people in the field can give an advantage of being able to shape the direction of the field, establish themselves as leading experts and gain a significant first mover advantage. However, as the field grows, the competition may increase and it may become harder to stand out and make significant contributions.

---

[86] *See* Marvin B. Lieberman & David B. Montgomery, *First-Movers Advantages*, 9 Strategic Mgmt. J. 41 (1988), https://doi.org/10.1002/smj.4250090706.

[87] *See* Amanda L. Vogel, Brooke A Stipelman, Kara L. Hall, Linda Nebeling, Daniel Stokols & Donna Sprujit-Metz, *Pioneering the Transdisciplinary Team Science Approach: Lessons Learned from National Cancer Institute Grantees*, 2 J. of Translational Med. & Epidemiology 1027 (2014), https://pubmed.ncbi.nlm.nih.gov/25554748/.

# Conclusion

In this article, we have presented a case for why the use of AI-assisted writing tools in a scholarly context falls within fair use and does not violate copyright. We have also provided a set of best practices for standard of care with regard to plagiarism, copyright, and fair use.

By following the best practices for standard of care, scholars and researchers can use AI-assisted writing tools in a responsible and legal manner, and ensure the quality and reliability of their work. As AI is likely to grow more capable in the coming years, it is appropriate–and likely inevitable–to begin the process of ethically integrating AI into scholarly writing activities. This article lays the legal and scholarly foundation for this process to unfold.

Overall, AI-assisted writing tools have the potential to revolutionize the way scholarly writing is done and bring new possibilities for the advancement of knowledge and the promotion of progress of science and the useful arts. The legal and ethical considerations should be taken into account and best practices for standard of care should be followed, but the use of AI in scholarly writing should be embraced and further researched.

# Acknowledgments

We thank Amanda McElfresh and Justin Shock for their invaluable assistance with this article. This material is based upon work supported by the National Science Foundation under Grant No. DUE-2121572.